\documentclass[aps,showpacs,graphicx]{revtex4}
\usepackage{graphicx}
\begin{document}

\title{Passively self-error-rejecting qubit transmission over a collective-noise channel\footnote{Published in Quantum Information and
Computation, Vol. 11, No. 11\&12 (2011) 0913 -- 0924}}
\author{ Fu-Guo Deng,$^{1,}$\footnote{Author to whom correspondence should be addressed. Email address: fgdeng@bnu.edu.cn} Xi-Han Li,$^{2}$ and Hong-Yu
Zhou$^{3}$}
\address{$^1$Department of Physics, Applied Optics Beijing Area Major
Laboratory, Beijing Normal University, Beijing 100875, China\\
$^2$ Department of Physics,  Chongqing University,
Chongqing 400044, China\\
$^3$College of Nuclear Science and Technology, Beijing Normal
University, Beijing 100875,  China}
\date{\today }

\begin{abstract}
We propose a passively self-error-rejecting single-qubit
transmission scheme for an arbitrary polarization state of a single
qubit over  a collective-noise channel, without resorting to
additional qubits and entanglement. By splitting a single qubit into
some wavepackets with some  Mach-Zehnder interferometers, we can
obtain an uncorrupted state with a success probability approaching
100\% via postselection in different time bins, independent of the
parameters of collective noise. It is simpler and more flexible than
the schemes utilizing decoherence-free subspace and those with
additional qubits. One can directly apply this scheme to almost all
quantum communication protocols based on single photons or entangled
photon systems against a collective noise.

\end{abstract}
 \maketitle

\section{introduction}

Quantum key distribution (QKD) supplies a secure way for two
parties, say the sender Alice and the receiver Bob, to generate a
shared key, provided that they initially share a short secret key
(for identity authentication) and that they possess an unprotected
quantum channel (an optical fiber). Different from classical
crypto-system in which the security of key depends on computation
difficulty with a limited computation power, the security of QKD
comes from the laws of quantum mechanics such as the uncertainty
relation (non-cloning theorem), the coherence of entangled systems,
quantum measurement, and so on. As an unknown quantum state cannot
be cloned, the vicious actions done by an eavesdropper, say Eve will
inevitably disturb the quantum system and leave a trace in the
outcomes obtained by the two authorized parties. Eve's action will
be detected by analyzing the error rate of samples chosen randomly.
Since Bennett and Brassard published the original QKD protocol
\cite{bb84} in 1984 (called BB84), QKD attracts a great deal of
attention
\cite{rmp,QKD1,QKD2,QKD3,QKD4,QKD55,QKD6,QKD7,QKD8,QKD9,QKD10,QKD11,QKD12,QKD13,QKD14,QKD15}
and has been proven unconditionally secure \cite{proof1,proof2}.
Recently, some groups demonstrated successfully long-distance
quantum cryptography \cite{qkdexper00,qkdexper1,qkdexper2,qkdexper3}
and its network
\cite{qkdnetworkexper0,qkdnetworkexper1,qkdnetworkexper2,qkdnetworkexper3,qkdnetworkexper4,qkdnetworkexper5}.

Implementations of practical QKD rely on either the polarization or
the differential phases of photons. Preventing Eve from
eavesdropping by disguising her action as noise with a better
quantum channel requires the two legitimate users to reduce the
influence of the noise in their quantum channels. Otherwise, they
can only distill a short shared key from a large raw string with
privacy amplification \cite{rmp}. When the noise in the quantum
channels is too large, secure key generation is impossible. For
overcoming the birefringence of the optical fiber which alters the
polarization state of photons, some QKD schemes are proposed with
Mach-Zehnder interferometers (MZIs) and a Faraday mirror which is
used to compensate polarization mode dispersion, such as the "plug
and play" QKD system \cite{plugplay1} and its modifications
\cite{plugplay2,plugplay3}. However, these two-way quantum
communication schemes are vulnerable to the Trojan horse attack
\cite{tha}. Also, it is not easy for the two legal users in quantum
communication to reduce the noise effect caused by the thermal
fluctuation, vibration, and the imperfection of the fiber. Recently,
some novel techniques are developed for protecting quantum
information transmission, such as decoherence-free subspaces (DFS)
\cite{Walton,B1,B2,QKD5}, error-correcting codes
\cite{book,onestep}, faithful qubit distribution
\cite{yamamoto,lixhimprove}, faithful qubit transmission
\cite{lixh}, error-rejecting codes \cite{correction}, and so on. In
DFS, a single logical qubit can be encoded in two physical qubits
\cite{Palma}, i.e., $\vert \bar{0}\rangle \rightarrow \vert H
V\rangle \equiv \vert H\rangle_{A_1}\vert V\rangle_{A_2}$, $\vert
\bar{1}\rangle \rightarrow  \vert VH\rangle \equiv\vert
V\rangle_{A_1} \vert H\rangle_{A_2}$. Here $\vert H\rangle$ and
$\vert V\rangle$ represent the horizontal polarization and the
vertical polarization, respectively. Usually, there is a time delay
$\Delta t$ between the qubit $A_1$ and the qubit $A_2$. This code
makes the logical qubits be immune to a collective-dephasing noise
which is described with a transformation \cite{Walton}: $\vert H
\rangle \rightarrow \vert H\rangle$, $\vert V \rangle \rightarrow
e^{i\phi}\vert V\rangle$ (the additional phase $\phi$ is unknown to
any one). Under this transformation, the states of two physical
qubits $\vert HV\rangle$, $\vert VH\rangle$, and
$\frac{1}{\sqrt{2}}(\vert HV\rangle \pm \vert VH\rangle)$ all are
immune to this collective-dephasing noise, and can be used for
quantum communication perfectly \cite{Walton}. Wang showed that DFS
can also used for QKD over a collective- random-unitary-noise
channel by checking parity and sacrificing a proportion of qubits
\cite{wangxb}. In error-correcting codes \cite{book}, at least five
entangled physical qubits are encoded for a single logical qubit
against the noise. In 2005, Yamamoto \emph{et al.} \cite{yamamoto}
introduced a good way for faithful qubit distribution with one
additional qubit against a collective noise. Their scheme can be
perfectly used for secure key generation with two quantum channels.
The proportion of uncorrupted qubits to those transmitted approaches
$1/8$ (it depends on the coefficients of the noise
\cite{lixhimprove}). More recently, a scheme \cite{lixh} for
faithful qubit transmission without additional qubits is proposed
with two quantum channels. Its proportion of uncorrupted qubits to
those transmitted approaches $1/2$ in a passive way. With some
delayers, the proportion can be improved to 1. In the
error-rejecting codes \cite{correction}, at least two fast
polarization modulators (Pockels cell), whose synchronization makes
it difficult to be implemented with current technology
\cite{coherent}, are employed \cite{lixh}. In  the quantum
error-rejection code protocol proposed by Wang \cite{wang2} against
bit-flipping errors with entanglement, the user should exploit a
parity-check tool to read out the qubit probabilistically.

In this paper, we introduce a scheme for passively
self-error-rejecting single-qubit transmission over a
collective-noise channel with a success probability approaching
100\%, several times of other schemes. For example, the success
probability in the present scheme is about eight times as that in
the scheme proposed by Yamamoto \emph{et al.} \cite{yamamoto} in a
passive way. Moreover, it is independent of the parameters of a
collective noise. Unlike other schemes
\cite{Walton,B1,B2,wangxb,wang2}, it does not require entanglement.
Different from  Yamamoto's scheme \cite{yamamoto}, the present
scheme needs no additional qubits and it also works for the
transmission of one photon in an entangled photon pair. Moreover,
our scheme works with one quantum channel, not two
\cite{yamamoto,lixh} or more, and its implement is based on some
simple optical devices. All these good features make it easy to
apply for almost all quantum communication protocols existing, such
as the quantum cryptography protocols based on single photons or
entangled photon systems.

\begin{center}
\begin{figure}[!h]
\includegraphics[width=10cm,angle=0]{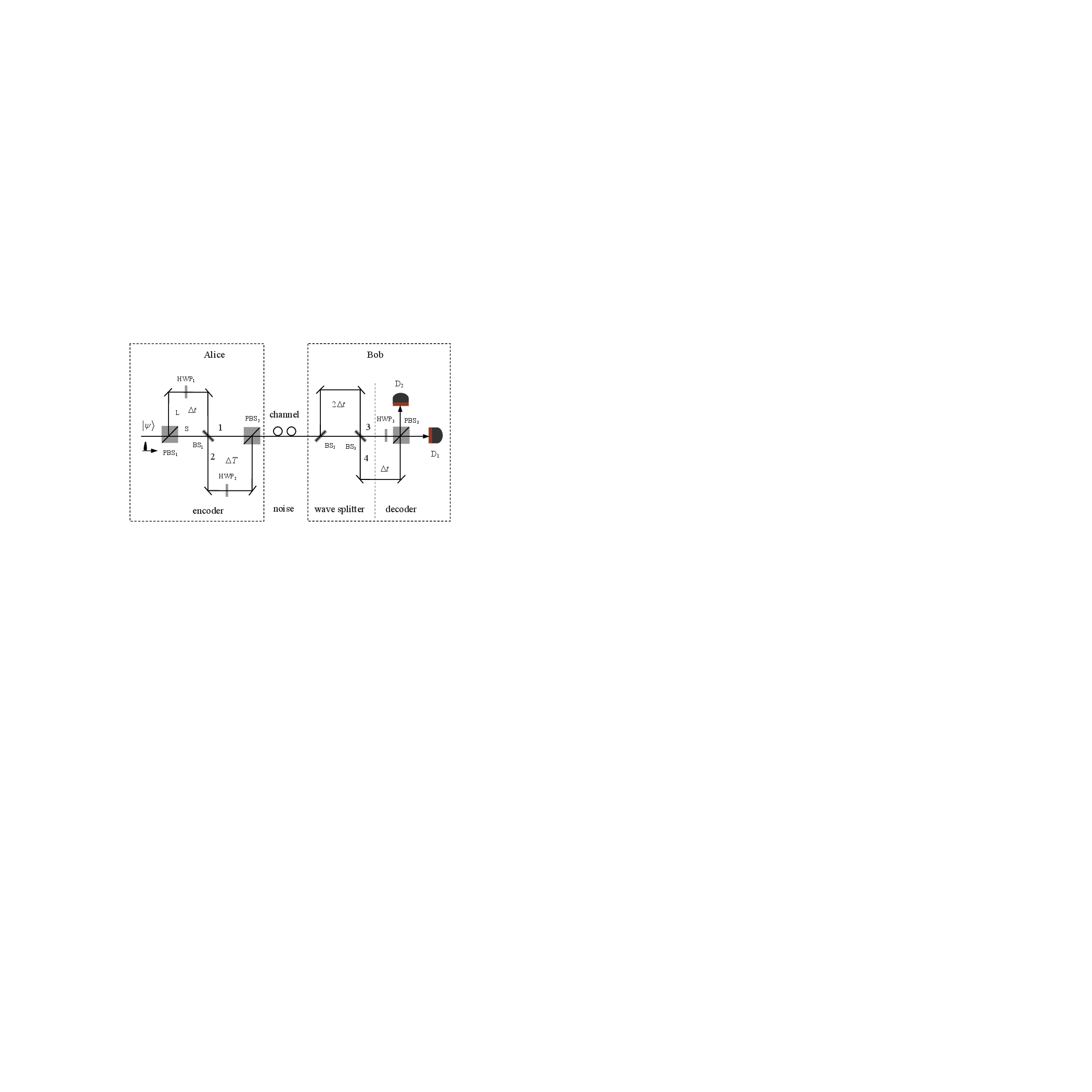}
\caption{Schematic representation of the present
self-error-rejecting single-qubit transmission scheme over a
collective-noise channel. PBS$_i$ ($i=1,2,3$), HWP, and BS$_i$
represent a polarizing beam splitter, a half wave plate, and a beam
splitter (50/50), respectively.  The intervals between the long path
and the short path in the two unbalanced Mach-Zehnder
interferometers are $\Delta t$ and $\Delta T$, respectively.}
\end{figure}
\end{center}

\section{Passively self-error-rejecting single-qubit transmission protocol}

The principle of our self-error-rejecting single-qubit transmission
scheme over a collective-noise channel is shown in Fig.1. It
comprises an encoder, a collective-noise channel, a wave splitter,
and a decoder.  The fluctuation in the collective-noise channel is
slow in time so that the alteration of the polarization is
considered to be the same over the sequence of several photons (or
wavepackets) \cite{yamamoto}. The encoder is made up of two
unbalanced Mach-Zehnder interferometers (MZIs) with different
intervals, i.e., $\Delta t$ ($\Delta t \equiv t_L - t_S$) and
$\Delta T$. A single qubit, whose original state is $\vert \psi
\rangle_0=\alpha \vert H\rangle + \beta \vert V\rangle$, is split
into two parts by the first polarizing beam splitter (PBS), which
transmits $\vert H\rangle$ and reflects $\vert V\rangle$. A half
wave plate (HWP) rotates the polarization of the photons in the path
$L$ by $90^{o}$, i.e., $\vert H\rangle \leftrightarrow \vert
V\rangle$. Before the first beam splitter (BS$_1$:50/50), the state
of the single qubit can be described as $\vert \psi \rangle_B=\alpha
\vert H\rangle_S +
 \beta \vert H\rangle_L \equiv \alpha \vert
H\rangle_{0} + \beta \vert H\rangle_{\Delta t}$. Therefore, the
single qubit before it enters the collective-noise channel is in the
state
\begin{eqnarray}
\vert \psi \rangle_C = \frac{1}{\sqrt{2}}(\alpha \vert H \rangle_0 +
i \beta \vert H \rangle_{\Delta t} + i\alpha \vert V \rangle_{\Delta
T} + \beta \vert V \rangle_{\Delta T + \Delta t})\equiv
\frac{1}{\sqrt{2}}(\vert \psi\rangle_H + \vert \psi\rangle_V),
\end{eqnarray}
where
\begin{eqnarray}
\vert \psi\rangle_H  &=& \alpha \vert H \rangle_0 + i \beta \vert H
\rangle_{\Delta t}, \label{hstate}\\
\vert \psi\rangle_V  &=&   i\alpha \vert V \rangle_{\Delta T} +
\beta \vert V \rangle_{\Delta T + \Delta t}.\label{vstate}
\end{eqnarray}
The complex coefficient $i$ comes from the phase shift aroused by
the BS$_1$ reflection (we assume that the surface of the BS$_1$ has
the phase shift $i$ between the wave packet reflected and that
transmitted), and the subscripts represent the signal time slots
arrived.

Suppose that the collective noise in an optical fiber transforms the
polarization states of a photon as
\begin{eqnarray}
\vert H \rangle & \rightarrow  & \delta_1 \vert H\rangle +
\eta_1\vert
V\rangle, \label{hrotation}\\
\vert V \rangle & \rightarrow &  \delta_2 \vert H\rangle + \eta_2
\vert V\rangle,\label{vrotation}
\end{eqnarray}
where
\begin{eqnarray}
\vert \delta_1 \vert^2  + \vert \eta_1\vert ^2=\vert \delta_2
\vert^2 + \vert \eta_2 \vert ^2 =1.
\end{eqnarray}
The four parameters $\delta_1$, $\eta_1$, $\delta_2$, and $\eta_2$
vary with the time $t$ slowly, which means that only the photons
transmitted close to each other suffer from the same noise. The
decoherence channels represented by the unitary transformations
shown in Eqs.(\ref{hrotation}) and (\ref{vrotation}) indicate that a
photon is in a pure  polarization state when it is emitted from the
noisy channel although it is rotated and its state is unknown to us
accurately (for a large number of single photons, we should use a
mixed state to describe the state of a photon statistically).

The states shown in Eqs.(\ref{hstate}) and (\ref{vstate}) have the
same form but different parameters, and so do the rotations arisen
from the noisy channels shown in Eqs.(\ref{hrotation}) and
(\ref{vrotation}). That is, Bob can distill an uncorrupted state
from the states $\vert \psi\rangle_H$ and $\vert \psi\rangle_V$ with
the same principle. We first discuss the principle of the decoder
for distilling an uncorrupted state from the state $\vert
\psi\rangle_H$ in detail as follows and then generalize it from the
state $\vert \psi\rangle_V$.

The rotation by the collective-noise channel on the state $\vert
\psi\rangle_H$ will transform it into the state $\vert
\psi'\rangle_H$, i.e.,
\begin{eqnarray}
\vert \psi \rangle_H \; ^{\;\underline{\; noise \;}}\rightarrow
\vert \psi'\rangle_H &=& \delta_1 (\alpha \vert H \rangle_0 + i
\beta \vert H \rangle_{\Delta t}) + \eta_1 (\alpha \vert V \rangle_0
+ i \beta \vert V \rangle_{\Delta
t}) \nonumber\\
& \equiv & \delta_1 [\alpha  + i\hat{D}(\Delta t) \beta ]\vert H
\rangle_0 + \eta_1 [\alpha  + i\hat{D}(\Delta t) \beta ]\vert V
\rangle_0\nonumber\\
& \equiv &   \delta_1 \vert \phi\rangle_H + \eta_1 \vert
\phi\rangle_V. \label{dgz1f}
\end{eqnarray}
Here
\begin{eqnarray}
\vert \phi\rangle_H &=& [\alpha  + i\hat{D}(\Delta t) \beta ]\vert H
\rangle_0,\nonumber\\
\vert \phi\rangle_V &=& [\alpha  + i\hat{D}(\Delta t) \beta ]\vert V
\rangle_0,
\end{eqnarray}
and $\hat{D}(\Delta t)$ is a time-delay operator. That is,
\begin{eqnarray}
\hat{D}(\Delta t)\vert \psi \rangle_0 &=& \vert \psi \rangle_{\Delta
t}, \nonumber\\
\hat{D}(\Delta t_1)\hat{D}(\Delta t_2) &=&\hat{D}(\Delta t_1+\Delta
t_2).
\end{eqnarray}
Bob uses a wave splitter and a decoder to distill an uncorrupted
state, shown in Fig.1. The time interval between the two paths of
the wave splitter is $2\Delta t$. The wave splitter and the decoder
has the same role for the states $\vert\phi\rangle_H$ and
$\vert\phi\rangle_V$ but different outports of the PBS$_3$. The
combination of the wave splitter and the decoder will complete the
transformation on the state  $\vert\phi\rangle_H$  as follows,
\begin{eqnarray}
\vert\phi\rangle_H \rightarrow \vert\phi'\rangle_H &=&
\{\hat{\sigma}_x+i\hat{D}(\Delta t)+ \hat{D}(2\Delta
t)[i^2\hat{\sigma}_x+i\hat{D}(\Delta
t)]\}\vert\phi\rangle_H \nonumber\\
&=&\{\hat{\sigma_x}\alpha + i\hat{D}(\Delta t)[ \hat{\sigma}_x \beta
+ \alpha]+ i^2 \hat{D}(2\Delta t)[\beta + \hat{\sigma}_x \alpha]
\nonumber\\
&&  + i\hat{D}(3\Delta t)[i^2\hat{\sigma}_x \beta + \alpha] + i^2
\hat{D}(4\Delta t)\beta\}\vert H \rangle_0\nonumber\\
& \equiv & \hat{K} \vert H \rangle_0, \label{decoderoperator}
\end{eqnarray}
where $\hat{K}$ is a quantum operator used for describing the
principle of the reconstruction of the unknown state $\vert
\psi\rangle_0$ and $\hat{\sigma}_x = \vert H \rangle\langle V \vert
+ \vert V \rangle\langle H \vert$ is a bit-flip operation. Bob can
get an uncorrupted state $\vert \psi\rangle_0$ from the outport
$D_2$ of the PBS$_3$ at the time slots $\Delta t$, $2\Delta t$, and
$3\Delta t$ with the unitary operations $I$, $\hat{\sigma}_x$,  and
$\hat{\sigma}_z$, respectively,  which takes place with the success
probability $3/4$, shown in Fig.2. That is,
\begin{eqnarray}
&\Delta t:& \;\;\;\; (\alpha + \hat{\sigma}_x \beta)\vert H \rangle=
\alpha\vert H \rangle+ \beta \vert V \rangle \; ^{\;\underline{\;\; I \;}}\rightarrow \alpha\vert H \rangle+ \beta \vert V \rangle,\nonumber\\
&2\Delta t&:\,\,\,\, (\hat{\sigma}_x \alpha + \beta)\vert H \rangle=
\alpha\vert V\rangle + \beta \vert H \rangle \;^{\;\underline{\;\hat{\sigma}_x}\;}\rightarrow \alpha\vert H \rangle+ \beta \vert V \rangle,\nonumber\\
&3\Delta t&:\,\,\,\, (\alpha + i^2\hat{\sigma}_x \beta)\vert H
\rangle=\alpha\vert H\rangle - \beta \vert V \rangle \;
^{\;\underline{\;\hat{\sigma}_z}\;}\rightarrow \alpha\vert H
\rangle+ \beta \vert V \rangle.
\end{eqnarray}
Here $I=\vert H\rangle \langle H\vert + \vert V\rangle\langle
V\vert$ and $\hat{\sigma}_z=\vert H\rangle \langle H\vert - \vert
V\rangle\langle V\vert$.

\begin{center}
\begin{figure}[!h]
\includegraphics[width=10cm,angle=0]{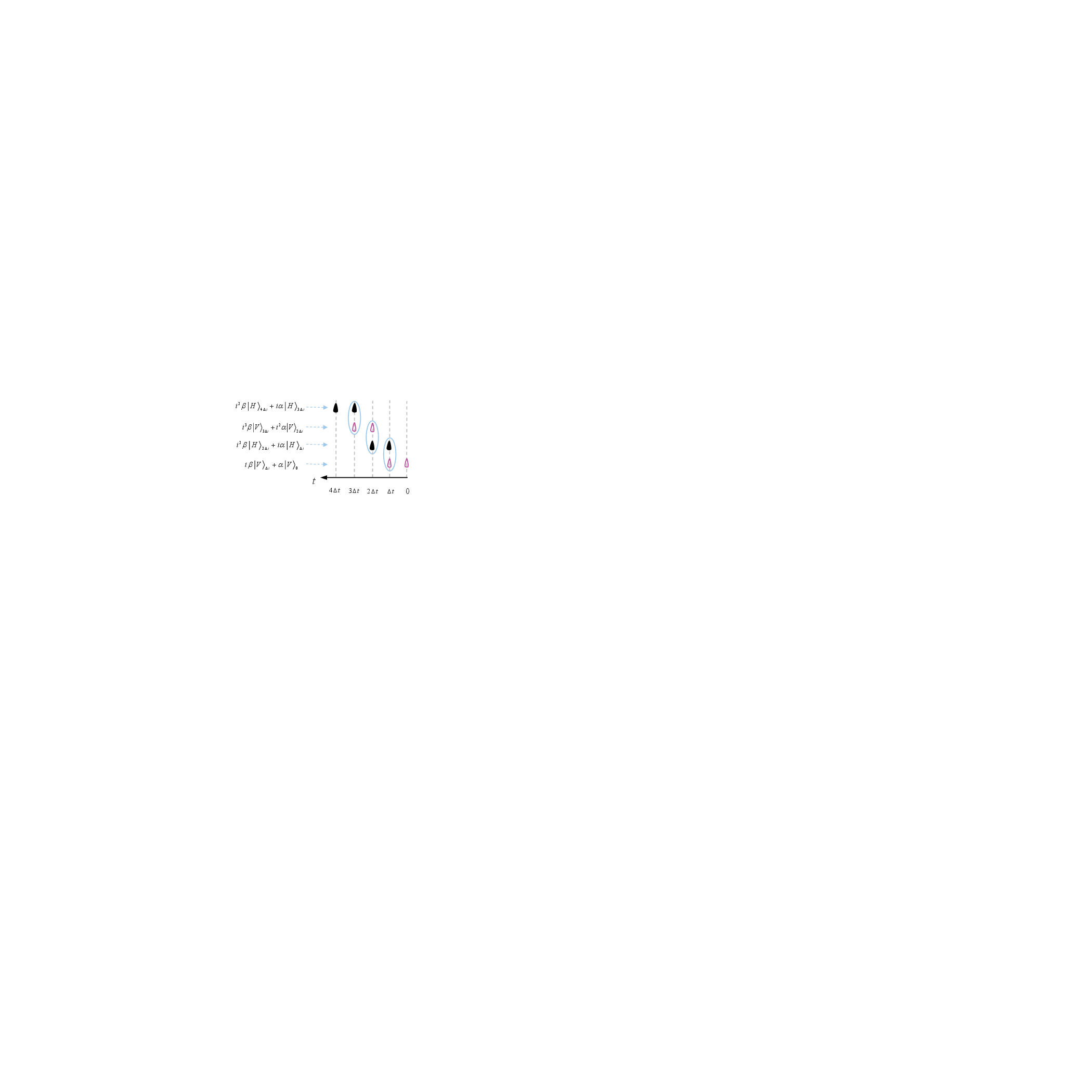}
\caption{Schematic representation for the reconstruction of the
original state $\vert \psi \rangle_0=\alpha \vert H\rangle +
\beta\vert V\rangle$ from the state $\vert \phi\rangle_H$. The two
wavepackets closed in an ellipse represent the fact that they will
emerge at the PBS$_3$ at the same time and interfere with each
other, which takes place with a success probability of $3/4$.}
\end{figure}
\end{center}

Bob can also distill an uncorrupted state from the state $\vert \phi
\rangle_V$ at the outport $D_1$ of the PBS$_3$, similar to the case
from the state $\vert \phi \rangle_H$. In detail, the combination of
the wave splitter and the decoder will complete the transformation
on the state  $\vert\phi\rangle_V$ as follows,
\begin{eqnarray}
\vert\phi\rangle_V \rightarrow \vert\phi'\rangle_V
&=&\{\hat{\sigma_x}\alpha + i\hat{D}(\Delta t)[ \hat{\sigma}_x\beta
+ \alpha]+ i^2 \hat{D}(2\Delta t)[\beta + \hat{\sigma}_x \alpha]
\nonumber\\
&&  + i\hat{D}(3\Delta t)[ i^2\hat{\sigma}_x \beta + \alpha]+i^2
\hat{D}(4\Delta t) \beta \}\vert V \rangle_0.
\end{eqnarray}
Bob can also get an uncorrupted state $\vert \psi\rangle_0$ from the
outport $D_1$ at the time slots $\Delta t$, $2\Delta t$, and
$3\Delta t$ with the unitary operations $\hat{\sigma}_x$, $I$,  and
$\hat{\sigma}_y$, respectively. That is,
\begin{eqnarray}
&\Delta t:& \;\;\;\; (\alpha + \hat{\sigma}_x \beta)\vert V \rangle=
\alpha \vert V \rangle + \beta \vert H \rangle  \;^{\;\underline{\;\hat{\sigma}_x\;}}\rightarrow \alpha\vert H \rangle+ \beta \vert V \rangle,\nonumber\\
&2\Delta t&:\,\,\,\, (\hat{\sigma}_x \alpha + \beta)\vert V \rangle=
\alpha\vert H\rangle + \beta \vert V \rangle \;^{\;\underline{\; I \;}}\rightarrow \alpha\vert H \rangle+ \beta \vert V \rangle,\nonumber\\
&3\Delta t&:\,\,\,\, (\alpha + i^2\hat{\sigma}_x \beta)\vert V
\rangle=\alpha\vert V\rangle - \beta \vert H \rangle \;
^{\;\underline{\;\hat{\sigma}_y\;}}\rightarrow \alpha\vert H
\rangle+ \beta \vert V \rangle.
\end{eqnarray}
Here  $-i\hat{\sigma}_y=\vert V\rangle \langle H\vert - \vert
H\rangle\langle V\vert$. In this way, Bob can get the uncorrupted
state $\vert \psi\rangle_0=\alpha\vert H \rangle+ \beta \vert V
\rangle$ from the states $\vert \phi\rangle_H$ and $\vert
\phi\rangle_V$ at the time slots $\Delta t$, $2\Delta t$, and
$3\Delta t$. At the time slots 0 and $4\Delta t$, Bob will lose the
useful information about the unknown state $\vert
\psi\rangle_0=\alpha\vert H \rangle+ \beta \vert V \rangle$ as he
can not distill the parameters $\alpha$ and $\beta$, which takes
place with the probability $1/4$.

We have discuss the principle that Bob distills an uncorrupted state
from the state $\vert \psi \rangle_H$ shown in Eq. (\ref{hstate}).
The principle that Bob distills an uncorrupted state from the state
$\vert \psi \rangle_V$ shown in Eq. (\ref{vstate}) is similar to
that from the state $\vert \psi \rangle_H$. The rotation by the
noisy channel on the state $\vert \psi\rangle_V$ will transform it
into the state $\vert \psi'\rangle_V$, i.e.,
\begin{eqnarray}
\vert \psi \rangle_V \; ^{\;\underline{\; noise \;}}\rightarrow
\vert \psi'\rangle_V &=& \delta_2 [i\alpha  + \hat{D}(\Delta t)
\beta ]\vert H \rangle_{\Delta T} + \eta_2 [i\alpha  +
\hat{D}(\Delta t) \beta ]\vert V
\rangle_{\Delta T}\nonumber\\
& \equiv &   \delta_2 \vert \Phi\rangle_H + \eta_2 \vert
\Phi\rangle_V. \label{dgz1f2}
\end{eqnarray}
Here
\begin{eqnarray}
\vert \Phi\rangle_H &=& [i\alpha  + \hat{D}(\Delta t) \beta ]\vert H
\rangle_{\Delta T},\nonumber\\
\vert \Phi\rangle_V &=& [i\alpha  + \hat{D}(\Delta t) \beta ]\vert V
\rangle_{\Delta T}.
\end{eqnarray}
The combination of the wave splitter and the decoder will complete
the transformation on the state  $\vert\Phi\rangle_H$  as follows,
\begin{eqnarray}
\vert\Phi\rangle_H \rightarrow \vert\Phi'\rangle_H
&=&\{i\hat{\sigma_x}\alpha + i^2\hat{D}(\Delta t)[ - \hat{\sigma}_x
\beta + \alpha]- i^3 \hat{D}(2\Delta t)[\beta -\hat{\sigma}_x
\alpha]
\nonumber\\
&&  + i^2\hat{D}(3\Delta t)[\hat{\sigma}_x \beta + \alpha]+i
\hat{D}(4\Delta t)\beta\}\vert H \rangle_{\Delta T}.
\end{eqnarray}
Bob can get an uncorrupted state $\vert \psi\rangle_0$ from the
outport $D_2$ at the time slots $\Delta T + \Delta t $, $\Delta T +
2\Delta t$, and $\Delta T + 3\Delta t$ with the unitary operations
$\hat{\sigma}_z$, $\hat{\sigma}_y$, and $I$, respectively. That is,
\begin{eqnarray}
&\Delta T + \Delta t:& \;\;\;\; (\alpha - \hat{\sigma}_x \beta)\vert
H \rangle=
\alpha\vert H \rangle - \beta \vert V \rangle \;^{\;\underline{\;\hat{\sigma}_z\;}}\rightarrow \alpha\vert H \rangle+ \beta \vert V \rangle,\nonumber\\
&\Delta T + 2\Delta t&:\,\,\,\, (\hat{\sigma}_x \alpha - \beta)\vert
H \rangle=
\alpha\vert V\rangle - \beta \vert H \rangle \;^{\;\underline{\;\hat{\sigma}_y\;}}\rightarrow \alpha\vert H \rangle+ \beta \vert V \rangle,\nonumber\\
&\Delta T + 3\Delta t&:\,\,\,\, (\alpha + \hat{\sigma}_x \beta)\vert
H \rangle=\alpha\vert H\rangle + \beta \vert V \rangle \;
^{\;\underline{\; I \;}}\rightarrow \alpha\vert H \rangle+ \beta
\vert V \rangle.
\end{eqnarray}

With the same way, Bob can also distill an uncorrupted state from
the wavepacket in the state $\vert\Phi\rangle_V$. In detail, the
combination of the wave splitter and the decoder will complete the
transformation on the state $\vert\Phi\rangle_V$ as follows,
\begin{eqnarray}
\vert\Phi\rangle_V \rightarrow \vert\Phi'\rangle_V
&=&\{i\hat{\sigma_x}\alpha + i^2\hat{D}(\Delta t)[ - \hat{\sigma}_x
\beta + \alpha]- i^3 \hat{D}(2\Delta t)[\beta -\hat{\sigma}_x
\alpha]
\nonumber\\
&&  + i^2\hat{D}(3\Delta t)[ \hat{\sigma}_x \beta + \alpha]+i
\hat{D}(4\Delta t)\beta  \}\vert V \rangle_{\Delta T}.
\end{eqnarray}
Bob can get an uncorrupted state $\vert \psi\rangle_0$ from the
outport $D_1$ at the time slots $\Delta T + \Delta t $, $\Delta T +
2\Delta t$, and $\Delta T + 3\Delta t$ with the unitary operations
$\hat{\sigma}_y$, $\hat{\sigma}_z$, and $\hat{\sigma}_x$,
respectively. That is,
\begin{eqnarray}
&\Delta T + \Delta t:& \;\;\;\; (\alpha - \hat{\sigma}_x \beta)\vert
V \rangle=
\alpha\vert V \rangle - \beta \vert H \rangle \;^{\;\underline{\; \hat{\sigma}_y \;}}\rightarrow \alpha\vert H \rangle+ \beta \vert V \rangle,\nonumber\\
&\Delta T + 2\Delta t&:\,\,\,\, (\hat{\sigma}_x \alpha - \beta)\vert
V \rangle=
\alpha\vert H\rangle - \beta \vert V \rangle \;^{\;\underline{\;\hat{\sigma}_z \;}}\rightarrow \alpha\vert H \rangle+ \beta \vert V \rangle,\nonumber\\
&\Delta T + 3\Delta t&:\,\,\,\, (\alpha + \hat{\sigma}_x \beta)\vert
V \rangle=\alpha\vert V\rangle + \beta \vert H \rangle
\;^{\;\underline{\; \hat{\sigma}_x \;}}\rightarrow  \alpha\vert H
\rangle+ \beta \vert V \rangle.
\end{eqnarray}
In this way, Bob can get the uncorrupted state $\vert
\psi\rangle_0=\alpha\vert H \rangle+ \beta \vert V \rangle$ from the
states $\vert \Phi\rangle_H$ and $\vert \Phi\rangle_V$ at the time
slots $\Delta T + \Delta t$, $\Delta T + 2\Delta t$, and $\Delta T +
3\Delta t$, which takes place with the success probability $3/4$.

In order to distinguish the uncorrupted state coming from the state
$\vert \psi\rangle_H$  or $\vert \psi\rangle_V$  when the photon
emits from the outports $D_1$ or $D_2$, $\Delta T$ should not be
zero. That is, Bob should make the wavepackets from $\vert
\psi\rangle_H$ and $\vert \psi\rangle_V$ attain the PBS$_3$ in
different time slots and they do not interfere with each other. For
simplification, Bob can choose $\Delta T=\frac{\Delta t}{2}$.

\begin{center}
\begin{figure}[!h]
\includegraphics[width=10cm,angle=0]{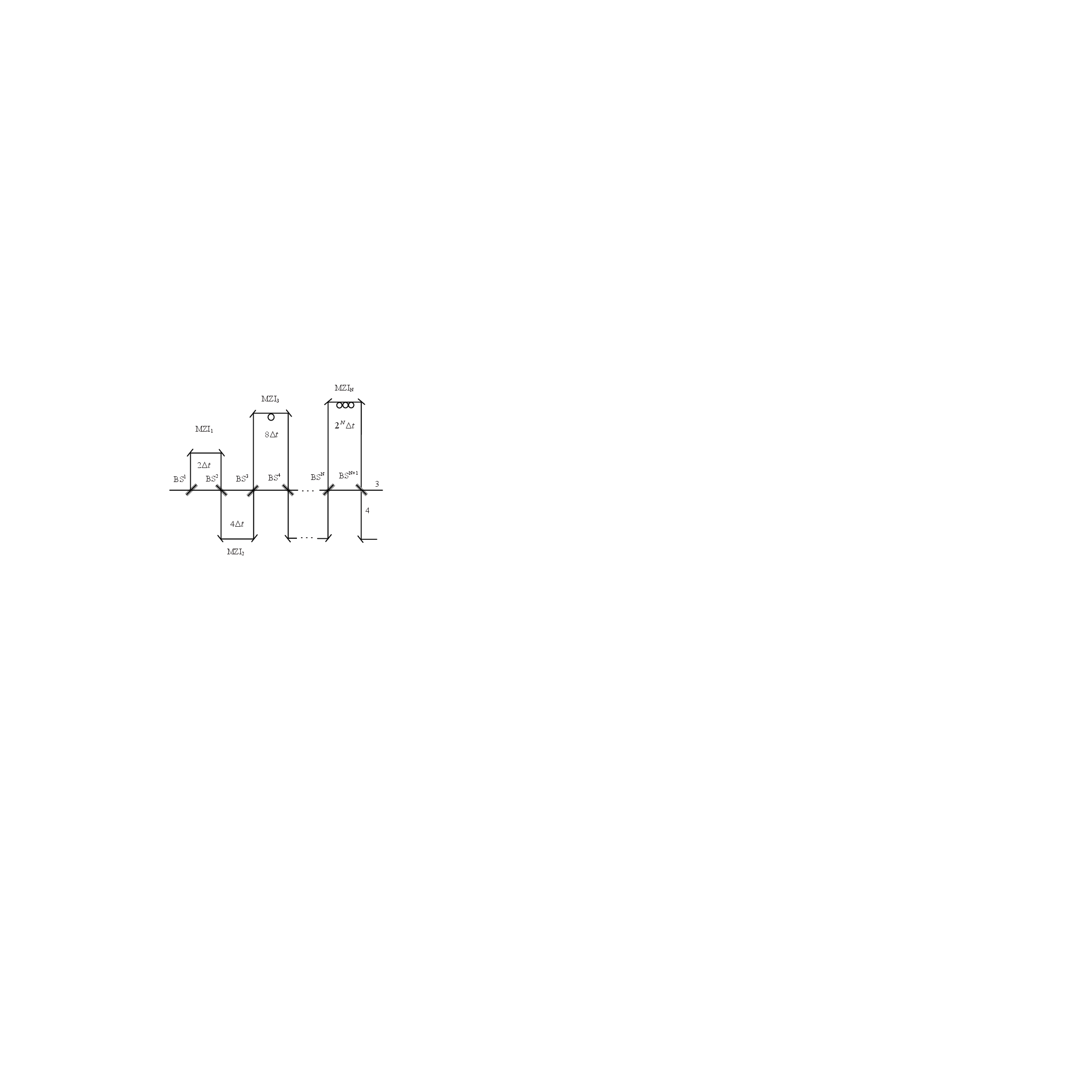}
\caption{A time-divisioned multiplexing for splitting a single qubit
into $2^{N}$ wavepackets with $N$ unbalanced MZIs.}
\end{figure}
\end{center}

\begin{center}
\begin{figure}[!h]
\includegraphics[width=10cm,angle=0]{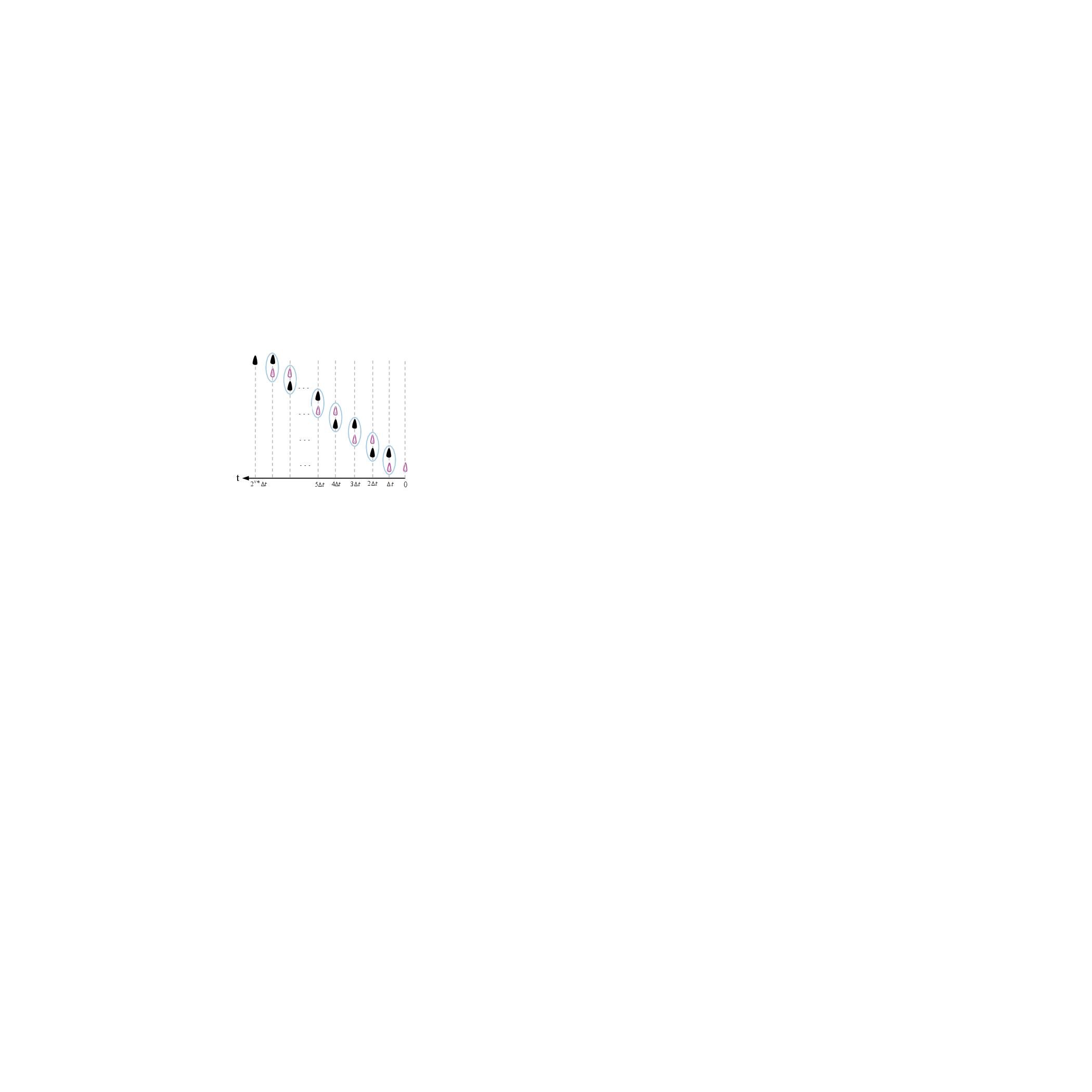}
\caption{Schematic representation for the reconstruction of the
original state $\vert \psi \rangle_0=\alpha \vert H\rangle +
\beta\vert V\rangle$ with $2^{N+2}$ wavepackets. The success
probability for obtaining an uncorrupted state is, in principle,
improved to be $P_S=\frac{2^{N+1}-1}{2^{N+1}}$.}
\end{figure}
\end{center}

From the discussion above, one can see that Bob can get an
uncorrupted state with the success probability 75\% if he exploits
the wave splitter shown in Fig.1 to split the wave pockets of the
single photon. Of course, Bob can improve the success probability by
using a time-divisioned multiplexing \cite{d1,d2,d3} to split the
single photon into more wavepackets, shown in Fig.3. In this time,
the quantum operator $\hat{K}$ in Eq. (\ref{decoderoperator}) will
be replaced with $\hat{K}'$. Here
\begin{eqnarray}
\hat{K}' \equiv &\{& a_0\hat{\sigma}_x \alpha \nonumber\\
&+& i\sum_{m=0}^{2^{N-1}- 1} a_{2m}
 \hat{D} [(2m + 1)\Delta t] (\alpha  +
 \hat{\sigma} _x \beta ) \nonumber\\
&+& \sum_{m=1}^{2^{N-1}}
\hat{D} (2m\Delta t) (a_{2m}\hat{\sigma}_x \alpha  - a_{2m - 2} \beta )\nonumber \\
&-& i\sum_{m=2^{N-1}}^{2^{N}- 1}a_{2m}
 \hat{D} [(2m + 1)\Delta t] (\alpha  -
 \hat{\sigma}_x \beta ) \nonumber\\
&+& \sum_{m=2^{N-1}+1}^{2^{N}-1}
\hat{D} (2m\Delta t) (a_{2m}\hat{\sigma}_x \alpha  + a_{2m - 2} \beta )\nonumber \\
&+&a_{2^{N + 1}  - 2}\hat{D} (2^{N + 1} \Delta t)\beta \},
\end{eqnarray}
where $a_j \in \{1, -1\}$ and can be determined when the number of
MZIs in the wave splitter $N$ is given. Also, Eq.
(\ref{decoderoperator}) will be transformed into
\begin{eqnarray}
\vert \phi\rangle_H \rightarrow \vert \phi''\rangle_H =\hat{K}'
\left| H \right\rangle _0.
\end{eqnarray}
With the same way as the case in which Bob chooses his wave splitter
shown in Fig.1, Bob can distill an uncorrupted state  $\vert \psi
\rangle_0=\alpha \vert H\rangle + \beta\vert V\rangle$ with the
success probability $P_S=\frac{2^{N+1}-1}{2^{N+1}}$, shown in Fig.4.
Moreover, this success probability is independent of the noise
parameters $\delta_1$, $\eta_1$,  $\delta_2$,   and $\eta_2$.

\section{discussion and summary}

In a practical application in quantum communication (such as QKD),
the qubit is measured immediately and no extra operations are
required for recovering the original state as the receiver can judge
the time when the qubit is detected and then he can compensate for
the effect of the extra operations by flipping the measured bit
value or not. That is, the present scheme is completely passive when
it is used as a part of a QKD protocol. Of course, this is the main
goal of the present scheme. Certainly, there are some other problems
when the present scheme is used in a practical quantum cryptography.
One is the effect of the channel losses and detection dark counts.
The other is the requirement that the delays by $\Delta t$ and
$2\Delta t$ should be done accurately, which means that the two
parties should possess some stable Mach-Zehnder interferometers. The
present scheme will suffer from the channel losses and detection
dark counts, the same as other faithful qubit transmission schemes
\cite{yamamoto,lixhimprove,lixh,correction} and quantum
communication protocols \cite{rmp}. In fact, the detection dark
counts will decease the key-generation rate as its effect equals to
lose a portion of the single photons transmitted over a noisy
channel. This is a general problem in quantum communication. On the
other hand, the channel losses has two effects. One is that it
decreases the key-generation rate if the photon is lost before it
arrives the side of the receiver. The other is that it will decrease
the success probability of the present faithful qubit transmission
scheme if only some wavepackets of the single photon are lost. In
the present scheme, the wavepackets of a single photon is so close
(not more than $\frac{3\Delta t}{2}$) that we can assume that the
wavepackets are lost or not as a whole system. Under this
assumption, the channel losses will decrease the key-generation rate
only, not the success probability. At present, it is not easy for us
to maintain the stabilization of a Mach-Zehnder interferometer for a
long time with only linear optical elements such as PBSs and BSs. On
one hand, this feature will improve the difficulty of the
implementation of the present scheme in a practical application. On
the other hand, the two parties in quantum communication can use
some reference signals to analyze periodically the  stability of the
Mach-Zehnder interferometers and compensate the fluctuation with
feedback. With the improvement of technology, the parties can also
use some interferometers with optical integrations in chips to
depress the fluctuation of time difference.

When the present scheme is used in some coherent quantum
communication protocols in which the qubits are not measured
immediately but stored, it does not work in a passive way. For
example, if the present scheme is used to distribute an entangled
photon pair for a quantum repeater (not for  generating a key
immediately  in long-distance QKD), the two photons with a high
fidelity will be stored for a period of time. At this time, the
parties should exploit some kinds of non-destructive quantum
measurements to detect the presence of the photons. It is not
necessary for the two parties to perform extra operations for
restoring the original state, just get the map of the correlation
between the unitary operations and the measured bit values obtained
later as they can also compensate for the effect of the extra
operations by flipping the measured bit value or not in the end of
quantum communication.

We have described a passively self-error-rejecting single-qubit
transmission scheme over a collective-noise channel. Compared with
the  scheme proposed by Yamamoto \emph{et al.} \cite{yamamoto} for
faithful qubit distribution assisted by one additional qubit and the
scheme without additional qubit \cite{lixh}, the present scheme has
some interesting features as follows: (1) The success probability
for obtaining an uncorrupted state $P_S=\frac{2^{N+1}-1}{2^{N+1}}$
approaches 100\% in principle if the number of wavepackets is large
enough (when $N=3$, $P_S=93.75\%$), which is about eight times of
that in the scheme introduced by Yamamoto \emph{et al.}
\cite{yamamoto}. At the aspect of success probability, the present
scheme is an optimal one. Of course, the bigger the number of the
wavepackets, the more time slots that Bob should pay for
reconstructing the original state, which maybe decrease the
key-generation rate in quantum communication. (2) The present scheme
does not require an additional qubit against a collective noise,
just the single qubit itself, which makes the present scheme have
some good applications in quantum communication. In detail, one can
easily apply this scheme to almost all quantum communication
protocols existing, such as the quantum cryptography protocols with
single photons or entanglement \cite{rmp}.  (3) The present scheme
requires only one quantum channel, not two or more
\cite{yamamoto,lixh}. (4) This scheme does not require fast
polarization modulators (Pockels cell) \cite{correction} when it is
used in quantum cryptography, i.e., it works in a completely passive
way for quantum cryptography with postselection. (5) It is easy to
implement this scheme with some simple optical devices in principle.
(6) The success probability does not depend on the extent of the
collective noise, i.e., it is independent of the noise parameters
($\delta_1$, $\eta_1$, $\delta_2$, and $\eta_2$), which is different
from those in Refs.\cite{yamamoto,wangxb}. As shown in
Eq.(\ref{dgz1f}) and Fig. 2, the wavepackets interfere with only
those with the same parameter of collective noise, and the success
probability for each part with the same noise parameter is
$P_S=\frac{2^{N+1}-1}{2^{N+1}}$. This good feature makes the present
scheme more efficient than other schemes \cite{yamamoto,lixh}. (7)
As the single qubit transmitted is in an arbitrary state $\vert \psi
\rangle_0=\alpha \vert H\rangle + \beta \vert V\rangle$, the present
scheme can also be used to accomplish the faithful transmission of
one particle in an entangled quantum system as an entangled pure
state $\alpha'\vert H\rangle_h\vert H\rangle_t + \beta'\vert
V\rangle_h\vert V\rangle_t$ can be rewritten as $\alpha''\vert
H\rangle_t + \beta'\vert V\rangle_t$ (here the subscript $h$ and $t$
represent the home particle and the traveling particle,
respectively).

In summary, we have present a passively self-error-rejecting
single-qubit transmission scheme for polarization states of photons,
which is immune to the collective noise in a quantum channel (an
optical fiber). The success probability for obtaining an uncorrupted
state, in principle, approaches 100\%  via postselection in
different time bins with some Mach-Zehnder interferometers,
independent of the parameters of collective noise, and the present
scheme can be implemented with some simple optical devices and
photon detectors in a completely passive way. The present scheme
does not employ an entangled state in DFS, and it does not resort to
additional qubits. One can directly apply this scheme to almost all
quantum communication protocols against a collective noise,
including the quantum cryptography protocols based on single photons
\cite{bb84,rmp} and those based on entangled photon systems
\cite{QKD1,QKD2,QKD3,QKD4}.

\section*{Acknowledgements} 

This work is is supported by the National Natural Science Foundation
of China under Grant Nos. 10974020 and 11174039, A Foundation for
the Author of National Excellent Doctoral Dissertation of P. R.
China under Grant No. 200723, the Beijing Natural Science Foundation
under Grant No. 1082008, and the Fundamental Research Funds for the
Central Universities.

\end{document}